\documentclass[10pt,a4paper]{article}
\usepackage[utf8]{inputenc}
\usepackage{amsmath}
\usepackage{amsfonts}
\usepackage{amssymb}
\usepackage{xfrac}
\usepackage{bm}
\usepackage{upgreek}
\usepackage{graphicx}
\usepackage[all]{nowidow}
\usepackage[left=2.5cm,right=2.5cm,top=2.5cm,bottom=2.5cm]{geometry}
\usepackage{cite}
\usepackage{dirtytalk}
\usepackage{float}
\usepackage[doublespacing]{setspace}
\usepackage{authblk}
\usepackage{setspace}
\usepackage{color,soul}
\usepackage{xcolor}
\usepackage{pdfpages}

\begin{document}

\title{Initiator-free photo-crosslinkable cellulose-based resists for fabricating submicron patterns via direct laser writing}

\author[1]{Maximilian Rothammer$^*$}
\author[2]{Dominic T. Meiers\thanks{contributed equally to this work}}
\author[1]{Maximilian Maier}
\author[2,3]{Georg von Freymann}
\author[1]{Cordt Zollfrank}

\affil[1]{Chair for Biogenic Polymers, Technische Universität München, 94315 Straubing, Germany}

\affil[2]{Physics Department and Research Center OPTIMAS, Technische Universit\"at Kaiserslautern, 67663 Kaiserslautern, Germany}

\affil[3]{Fraunhofer Institute for Industrial Mathematics ITWM, 67663 Kaiserslautern, Germany}

\maketitle
\newpage

\onehalfspacing

%%%%%%%%%%%%%%%%%%% abstract %%%%%%%%%%%%%%%%

\begin{abstract}
Novel bifunctional cellulose diacetate derivatives were synthesized in order to achieve bio-based photoresists, which can be structured by two-photon absorption via direct laser writing (DLW) without the need to use a photoinitiator. Therefore, cellulose diacetate is functionalized with thiol moieties and olefinic or methacrylic side groups enabling thiol-conjugated crosslinking. These cellulose derivatives are also photo-crosslinkable via UV irradiation ($\lambda$ = 254\,nm and 365\,nm) without using an initiator. 
\end{abstract}

%%%%%%%%%%%%%%%%%%%%%%%%%%  body  %%%%%%%%%%%%%%%%%%%%%%%%%%
\section{Introduction}

Direct laser writing via two-photon absorption enables the fabrication of three-dimensional architectures on the nanometer scale \cite{Malinauskas2011,Hohmann2015}. The polymerization of the photoresist is normally based on the simultaneous absorption of two photons of light in order to excite an photoinitiator molecule. This initiates the polymerization of a photosensitive material via radical polymerization. \cite{Delaney2021} Indeed, the need to use a photoinitiator may introduce several disadvantages such as the degradation of photoinitiators during long-term exposure to sunlight, which can lead to a yellowing tint of transparent polymer structures. Another drawback is a potential migration of initiator fragments out of the highly-defined spot (voxel) of the laser \cite{Delaney2021}, which might cause undesired polymerization of the surrounding resist. In additon, photoinitiators can be toxic and expensive. \cite{Scherzer2008}
Since DLW can be employed for the generation of complex hierarchical 3D photonic architectures \cite{Delaney2021}, we aim at the fabrication of bioinspired photonic and disordered patterns with photoinitiator-free bio-based photoresists.

The photopolymerization of acrylate-based photoresists proceeds via a radical chain-growth mechanism leading to insoluble polymer networks. Such insoluble polymer networks can also be achieved by step-growth processes such as radical thiol-ene reaction, where thiol and alkenes form thioether bonds \cite{Bagheri2019}.

Light induced thiol-ene crosslinking gained significant interest in the field of bioprinting and biofabrication due to its fast curing mechanism and biocompatibility of the materials \cite{Lee2020}. Thiol-ene reactions are not susceptible to oxygen inhibition and require low radical concentrations for photochemical initiation. Various hydrogel systems based on a norbornene-functionalized biopolymer and a multi-thiol crosslinker have been developed \cite{Gramlich2013,Ooi2018}. Three-dimensional microstructures were also fabricated employing DLW process via radical thiol-ene polymerization with a resin system based on a two-photon photoinitiator and multifunctional thiols and olefins \cite{Quick2013}. Vegetable oil based thiol-ene resins were also applied for 3D micro- and nanolithography \cite{Grauzeliene2021}. 
Generally photopolymerizations require the presence of photoinitator molecules, which generate reactive radical species upon absorption. Thiol-ene polymerization can be initiated without the use of any photoinitiator \cite{Cramer2002}. Therefore limiting some of its drawbacks like degradation-induced yellowing, toxicity and negative influences on curing depths.

As already described by Rothammer \textit{et al.} cellulose diacetate (CDA) was used as starting material to produce methacrylated CDA. Its solubility in organic solvents enables functionalization with photoreactive groups and ensures the required solubility to prepare photoresists \cite{Rothammer2018}. Our approach is to extend this previous work with new functionalities to synthesize a cellulose-based material which is photo-crosslinkable without the addition of photoinitiator. Therefore, we modify CDA with different functionalities (methacrylate, 4\hbox{-}pentenoate, thiol moiety) resulting in two derivatives, CDA-SH methacrylate and CDA-SH 4\hbox{-}pentenoate. These bifunctional cellulose derivatives are soluble in several organic solvents and are photo-crosslinkable by UV irradiation or employing a DLW process via two-photon absorption without the presence of a photoinitator. Thus, this biopolymers expand the class of photoresists based on renewable resources.

\section{Methods}

\subsection{Syntheses and characterization of the cellulose derivatives}

\subsubsection{Materials}
Cellulose diacetate (average M$_{n}$ \textasciitilde  30000 by SEC, 39.8\,wt\% acetyl), 3,3’\hbox{-}dithiodipropionic acid, 4\hbox{-}pentenoic acid and methacrylic anhydride were purchased from Sigma-Aldrich (USA). Dichloromethane, tetrahydrofurane, n\hbox{-}heptane, ethanol and chloroform were obtained from VWR Chemicals (France). \textit{N,N}\hbox{-}Dicyclohexylcarbodiimide (DCC), benzyl alcohol and 1,4\hbox{-}dithiothreitol (DTT) were acquired from Alfa Aesar (China). 4\hbox{-}(Dimethylamino)pyridine (DMAP) was purchased from Roth (Germany). Triethylamine (TEA) was obtained from Arcos Organics (France). All chemicals were used as received without further purification.

\subsubsection{Syntheses}
The production of bifunctional CDA derivatives containing reactive thiol and olefinic or methacrylic groups consists of a three step synthesis. For the introduction of reactive thiol groups in the backbone of CDA, a thiolation strategy described in literature is applied \cite{Andren2017}. This strategy is combined with esterification reactions based on DCC-coupling introducing methacrylate or 4\hbox{-}pentenoate functionalities in the CDA backbone. Detailed synthesis procedures are described in the supplemental document.

\subsubsection{UV-induced crosslinking}
For the preparation of photoresist, 50.0\,mg bifunctional CDA derivative are dissolved in 500\,µL acetone overnight. The photoresist is irradiated with wavelengths of 254\,nm and 365\,nm for 20\,minutes whereas the sample is 5 cm away from the UV light source. UV-induced crosslinking is conducted with a 8\,Watt UV-lamp (Herolab, Germany).

\subsubsection{Fourier transform infrared spectroscopy}
Fourier transform infrared (FTIR) spectra of CDA derivatives were recorded in MIR Frontier spectrometer (Perkin Elmer, USA) in attenuated total reflection mode. The spectra were recorded between 4000 and 500\,cm$^{-1}$ with a resolution of 4\,cm$^{-1}$ and 16\,scans.

\subsubsection{Nuclear magnetic resonance spectroscopy}
Nuclear magnetic resonance (NMR) spectra were acquired on a Jeol ECS-400 NMR (Japan). The evaluation was performed by MNova (Mestrelab Research, Spain). Samples were analyzed in CDCl$_{3}$ and d$^{6}$-acetone solutions at 25 °C. $^{13}$C NMR spectra were recorded with at least 2048\,scans whereas $^{1}$H-NMR spectra were recorded with 32\,scans. 

\subsubsection{Elemental analysis}
The elemental composition of the unmodified and modified CDA was measured by an EuroEA-Elemental Analyser (Eurovector, Italy). 1–3\,mg powder in tin crucibles was analyzed. This enabled the determination of the degree of substitution of CDA.

\subsection{Preparation of photoresists and direct laser writing}
To mix the photoresists a small amount of CDA-SH methacrylate and CDA-SH 4\hbox{-}pentenoate, respectively, is weighed and then dimethyl formamide (DMF, VWR International GmbH, Germany) is added until the specified concentration is reached. In the case of CDA-SH 4-pentenoate the resists are stirred over night using a magnetic stirrer. Using CDA-SH methacrylate the resists are allowed to rest over night without stirring. Eventually, the resists are filled into a reaction vial and separated for 20\,minutes in a centrifuge to eliminate unsolved particles from the resists used for DLW. After separation the shelf-life of the resists is about one day, hence they are applied for DLW within the same day. Afterwards, the resists start to polymerize by themselves since a thermally activated polymerization occurs over time. However, stored at -18\,°C the derivatives used for mixing the resists possess a shelf-life of a couple of months.

The structures are fabricated with a Photonic Professional GT (Nanoscribe GmbH, Germany) using a wavelength of 780\,nm with a laser power of 57.8\,mW (= 100\%). Using a syringe the photoresists are extracted from the vial and a drop of resist is positioned on top of a 170\,µm thick glass coverslip (Gerhard Menzel GmbH, Germany). Coating the top side of the substrate with a thin layer of aluminum oxide using atomic layer deposition (R-200 Standard, Picosun Oy., Finland) allows for a better detection of the glass-resist interface. The sample holder is prepared to enable sandwiching the resist between two coverslips by placing a second, uncoated coverslip slightly above the first one. This minimizes the exposure of the resist to atmosphere during a print job. The laser beam is focused into the resist through the substrate, applying a drop of immersion oil (Immersol 518F, Carl Zeiss AG, Germany) between the objective (63$\times$, NA = 1.4) and the bottom of the substrate. Once the printing is done, the samples are transferred to a beaker containing acetone for 30\,min followed by one containing isopropyl for 10\,minutes to wash the unpolymerized resist away in the case of CDA-SH 4-pentenoate based resists. For CDA-SH methacrylate based resists the developing routine consists of a 10\,min bath in DMF followed by 75\,min in acetone and 5\,min in isopropyl. Finally the samples are carefully blown dry using a nitrogen pistol.

\section{Results and discussion}

\subsection{Chemical functionalization of cellulose diacetate}

\begin{figure}
    \centering\includegraphics[width=15cm]{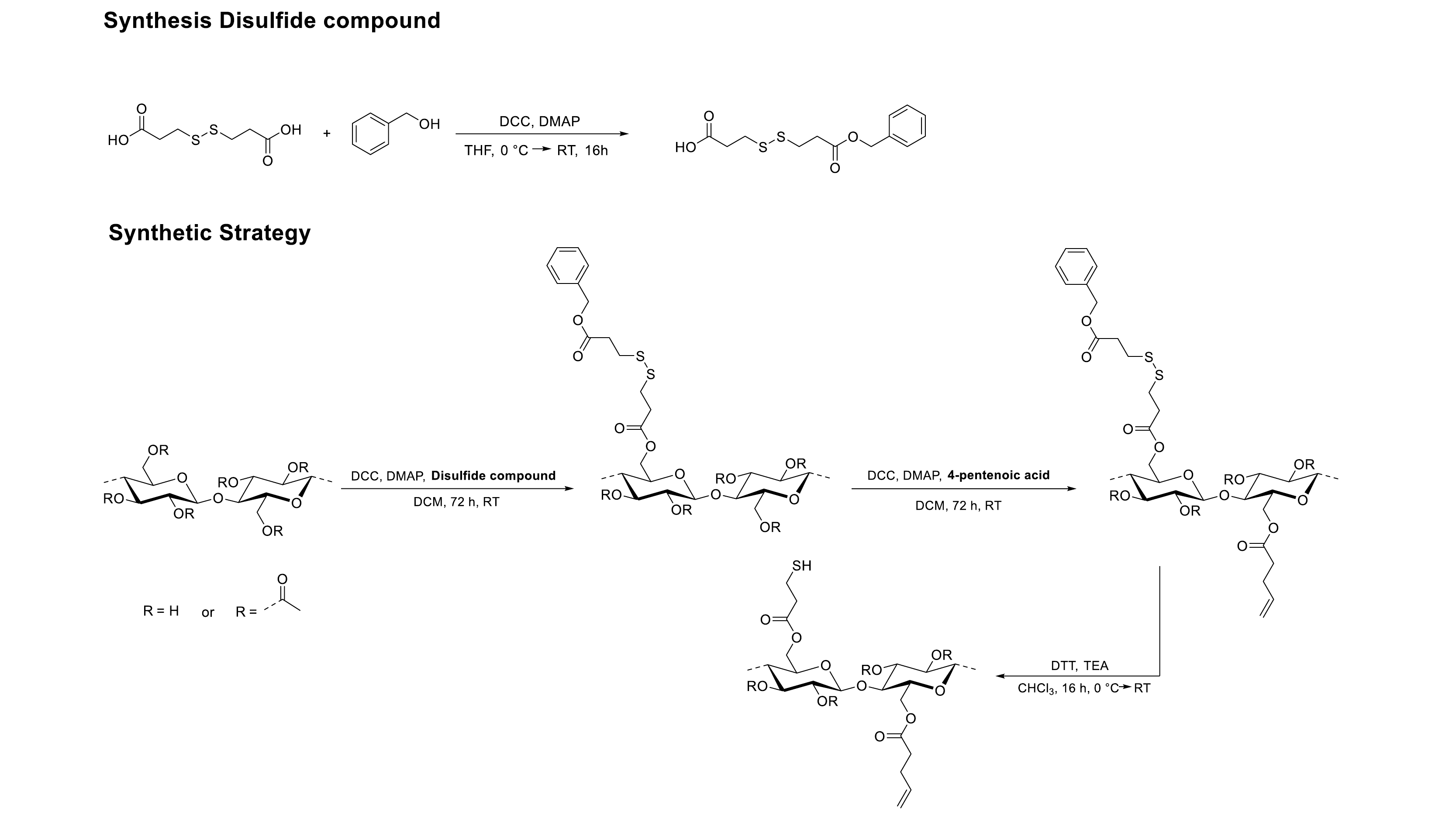}
    \caption{ Schematic representation of the synthetic strategy. R = H, acetyl, disulfide substituent, olefin or thiol substituent depending on the DS}
    \label{fig:Schematic}
\end{figure}

In order to employ thiol-conjugated reactions for photo-crosslinking, it is necessary to functionalize the polysaccharide backbone with thiol as well as olefinic or methacrylic functionalities. DCC as a coupling agent and DMAP as catalyst are used for linkage of methacrylate and 4\hbox{-}pentenoate to the CDA backbone \cite{Hasegawa2009,Nielsen2010}. For the introduction of reactive thiol-groups, a thiolation strategy centered around a monobenzylated asymmetric disulfide described in literature is applied \cite{Andren2017}. 3,3'-dithiodipropionic acid was used to synthesize an asymmetric disulfide with one dormant end and one reactive carboxylic group. In the first two reaction steps the disulfide compound and olefinic or methacrylic groups are introduced in the cellulose backbone by exploiting DCC-coupling. Finally disulfide cleavage using the reducing agent dithiothreitol results in thiol activation. A schematic depiction of the synthesis of CDA-SH 4\hbox{-}pentenoate including all synthesis steps is displayed in Fig. \ref{fig:Schematic}.

\begin{figure}
    \centering\includegraphics[width=15cm]{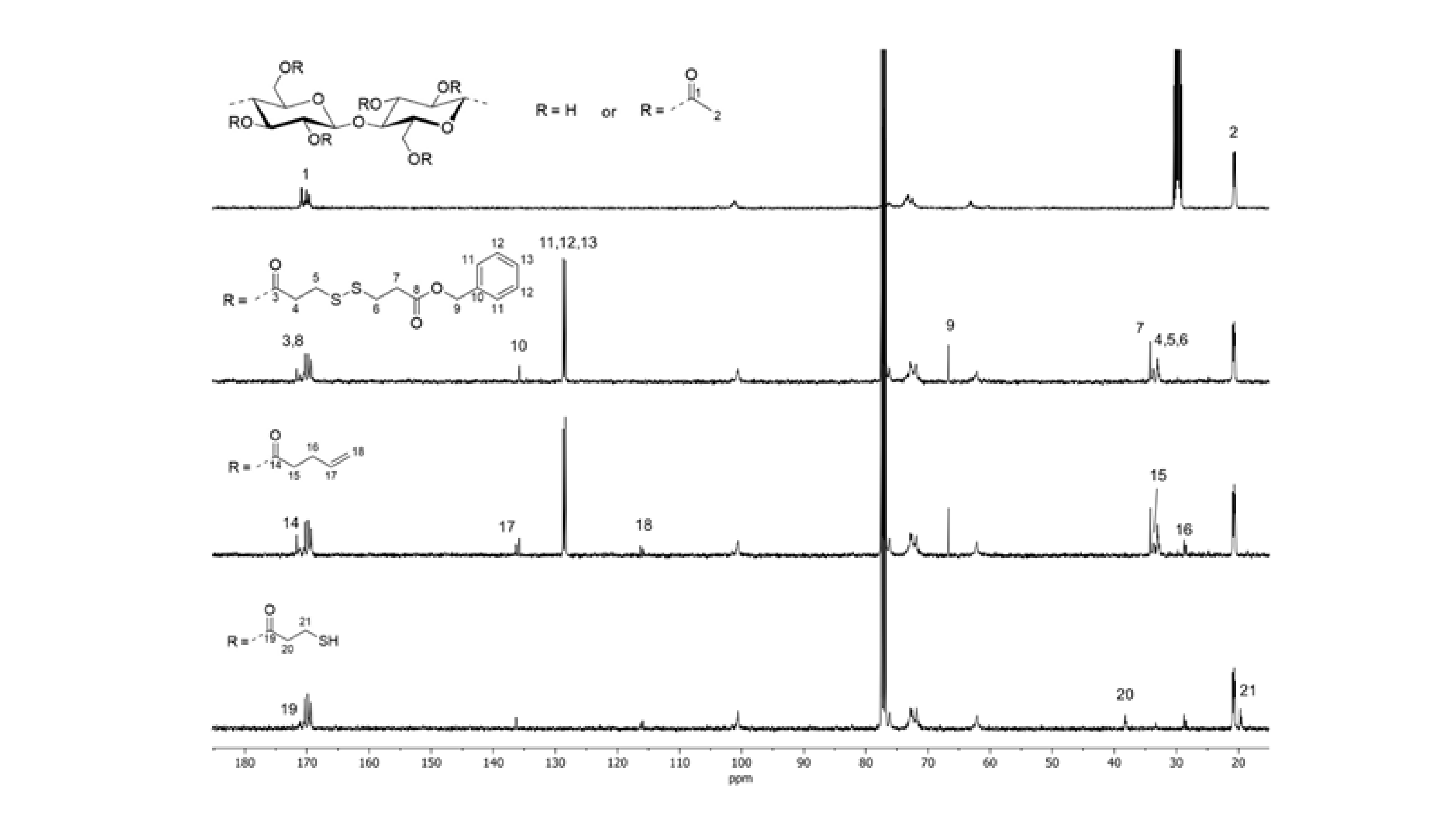}
    \caption{$^{13}$C NMR spectra of CDA, dp-CDA, dp-CDA 4\hbox{-}pentenoate, CDA-SH 4\hbox{-}pentenoate with assignment of relevant peaks.}
    \label{fig:NMR}
\end{figure}

The evaluation of the $^{13}$C NMR spectra confirms the successful functionalization of CDA and the existence of the reactive groups in the bifunctional cellulose derivatives. In Fig. \ref{fig:NMR} the NMR spectra of CDA, disulfide protected CDA (dp-CDA), dp-CDA modified with 4\hbox{-}pentenoic acid, and bifunctional CDA are depicted. The characteristic signals of CDA appear at 170\,ppm, which are assigned to the acetate esters, and at 20\,ppm, which is associated with the methyl group of the acetate \cite{Rothammer2018}. The signals of the C1–C6 atoms of the pyranering carbons are detected in the range from 100 to 62\,ppm. After introduction of the disulfide protection group new peaks appear at 172.1, 137.5, 129.1, 66.9 and 34.0 – 33.6\,ppm. All these peaks can be assigned to the monobenzylated disulfide compound. Functionalization with 4\hbox{-}pentenoic acid leads to peaks at 137.8 and 115.9 ppm, which can be assigned to carbon atoms of the double bond. Moreover, the peaks at 33.0 and 28.4 – 28.1\,ppm derive from the aliphatic carbon atoms of the ester substituent. After the reductive cleavage of the disulfide compound the earlier described peaks attributed to the aromatic carbon atoms and the peaks at 66.9 and 34.0 – 33.6\,ppm disappear. In fact two new signals arise with chemical shifts at 38.2 and 19.6\,ppm corresponding to the methylene protons. The carbon atom in ultimate proximity to the sulfur atom undergoes a shielding effect of the activated thiol group and generates a signal with a lower chemical shift. The signal of the carbon atom next to the ester group shifts to a higher ppm-value. The $^{13}$C NMR spectra confirm the successful functionalization of CDA and the production of multifunctional cellulose derivatives. The $^{13}$C NMR-Spectrum of CDA is recorded in acetone\hbox{-}d$^{6}$, whereas all other spectra are measured in CDCl$_{3}$. 

\begin{figure}
    \centering\includegraphics[width=11cm]{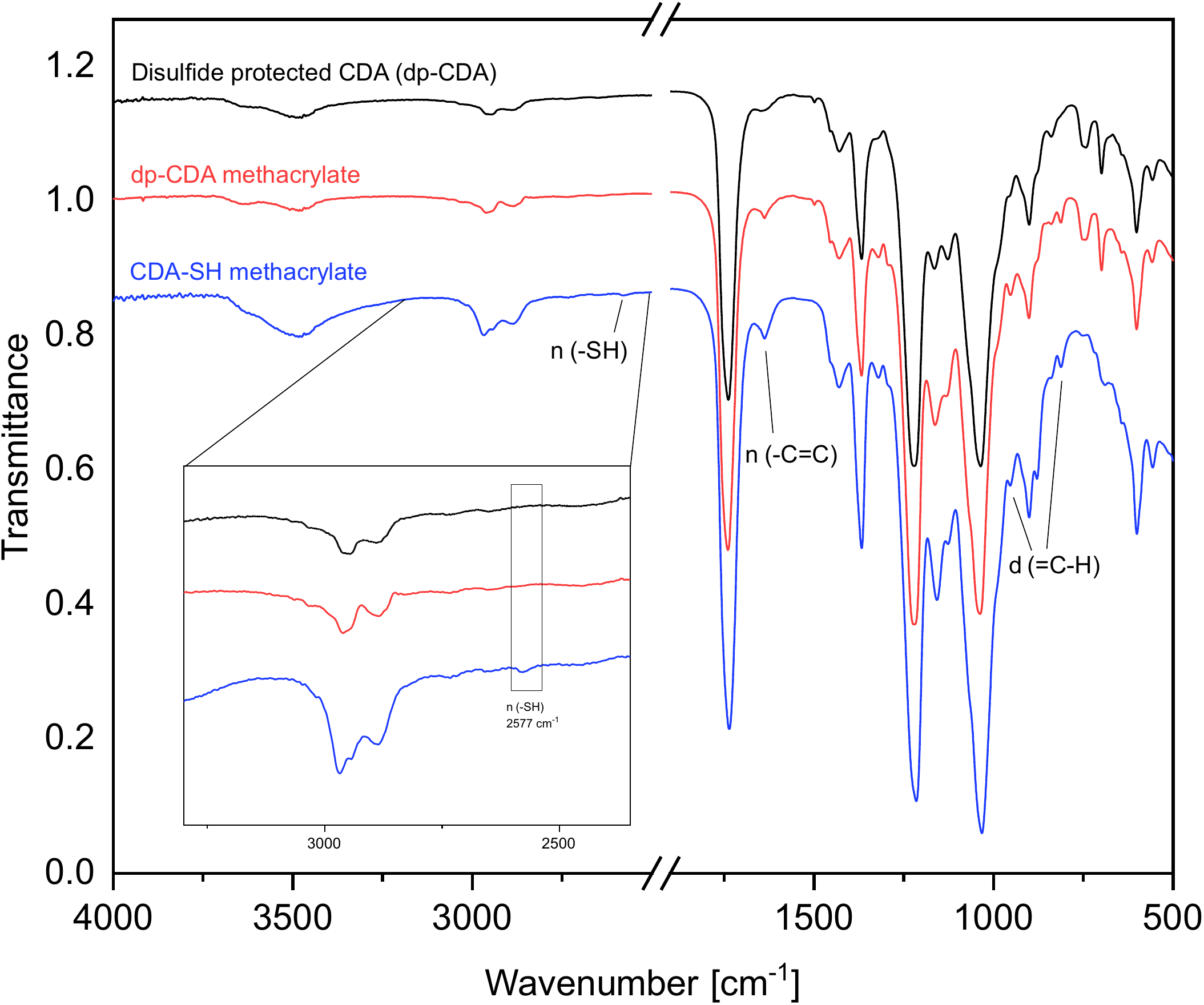}
    \caption{FTIR spectra of dp-CDA, dp-CDA methacrylate and CDA-SH methacrylate. The spectra of dp-CDA methacrylate and CDA-SH methacrylate are vertically shifted for better readability.}
    \label{fig:FTIR}
\end{figure}

Fourier transform infrared spectroscopy (FTIR) analyses are performed on the products and the different intermediates. This allows for the verification of the introduction of the disulfide protection group, the esterification with methacrylic anhydride or 4-pentenoic acid, the deprotection and consecutive thiol activation. The most significant FTIR absorption bands for these different CDA derivatives are represented in Fig. \ref{fig:FTIR}. The characteristic FTIR absorption bands of CDA are widely described in literature \cite{Barud2008,Kono2015}. After disulfide protection of CDA the characteristic bands of the benzyl protection groups 1499\,cm$^{-1}$, 747\,cm$^{-1}$ and 699\,cm$^{-1}$ appear. After methacrylation of CDA, new absorption bands arise at 1634\,cm$^{-1}$, 950\,cm$^{-1}$ and 811\,cm$^{-1}$. These bands correspond to the vibrations of alkene C=C stretching vibration and the C=CH$_{2}$ out of plane deformation vibration of the methacrylate group \cite{Rothammer2018}. Reduction of the protective disulfide leads to cleavage of the benzyl group and hence, to thiol activation. Therefore, the characteristic benzyl bands disappear and a band at 2577\,cm$^{-1}$ with low intensity corresponding to the S-H stretching vibration arises \cite{Huang2014}.

The degrees of substitution (DS) of the CDA derivatives were evaluated by $^{1}$H NMR measurements after functionalization. From the ratio between integrals of characteristic proton signals of the substituents and the integral of the seven protons of the anhydroglucose unit, the DS of the different intermediates could be determined. Characteristic proton signals for the disulfide compound are the five protons of the aromatic ring and the signals of vinyl protons for the methacrylate and 4\hbox{-}pentenoate functionalities. The DS after disulfide protection of CDA is determined between 0.2 and 0.3. The DS of olefinic functionalities is determined to 0.2 for methacrylic groups as well as for 4\hbox{-}pentenoate groups. After deprotection and therefore disulfide activation the DS of thiol groups can be also determined to a value of 0.2. In summary, we find equimolar distributions of functionalities for both bifunctional CDA derivatives, which provide suitable conditions for thiol conjugated crosslinking.

UV-induced photocrosslinking of CDA-SH methacrylate can be observed macroscopically or followed directly in the FTIR-spectra by the decrease or the disappearance of the methacrylate carbon double bond $\nu$(C=C) intensities at 1634\,cm$^{-1}$, 950\,cm$^{-1}$ and 811\,cm$^{-1}$. Due to its low intensity, a significant decrease of the thiol band at 2577\,cm$^{-1}$ cannot be seen in the FTIR-spectrum. But rapid crosslinking despite the absence of a photoinitator gives reason to assume an involvement of thiol-conjugated reactions in the photo-crosslinking process.

\subsection{Structure fabrication via direct laser writing}

Since the added thiol functionalities are designed to allow for photo-crosslinking the cellulose derivatives without any additive, searching for a suitable photoinitiator is not required. However, with the choice of the right solvent as well as a proper concentration, there are still free design parameter. 

While DLW is in principle a fast fabrication technology, printing structures with fine details or relatively large dimensions can quickly take a couple of hours or even longer. Therefore, acetone and chloroform, both capable to dissolve the cellulose derivatives are discarded as solvents for a DLW-suitable resist since they are highly volatile. Thus, the chemical and physical properties of the resist cannot be maintained during a print job, hindering reliability and reproducibility. In contrast, dimethyl sulfoxide (DMSO) and dimethyl formamide (DMF), which also dissolve the cellulose derivatives are barely volatile enabling a stable system. Both solvents are tested in respect of photoresist formulation revealing constantly better results when DMF is used. Indeed, exposing already polymerized structures to DMSO shows a degeneration of the structures revealing that DMSO attacks the polymer. For DMF no such effect can be observed, making it the solvent of choice.

Finding a proper concentration of the cellulose derivatives in the solvent is always a trade-off. Providing low amounts of cellulose leads to a rather inviscid resist. Hence, a drop of resist placed on the substrate tends to flow apart over time limiting the maximum height as well as introducing unwanted dynamics within the drop. On the other hand, using too much cellulose leads to a highly viscose resist, which can not be placed on the substrate without introducing unintended air bubbles in the drop. In addition, the viscosity also has impact on the diffusion of the radicals influencing the achievable resolution and feature size \cite{Waller2016}. Testing different resists it is found that for CDA-SH 4-pentenoate concentrations between 100\,g/l and 140\,g/l are a good trade-off with respect to handling of the resist as well as enabling the fabrication of structures via two-photon polymerization. In contrast for CDA-SH methacrylate the best results are obtained for a noticeably lower concentration of around 50\,g/l.

\begin{figure}
    \centering\includegraphics[width=15cm]{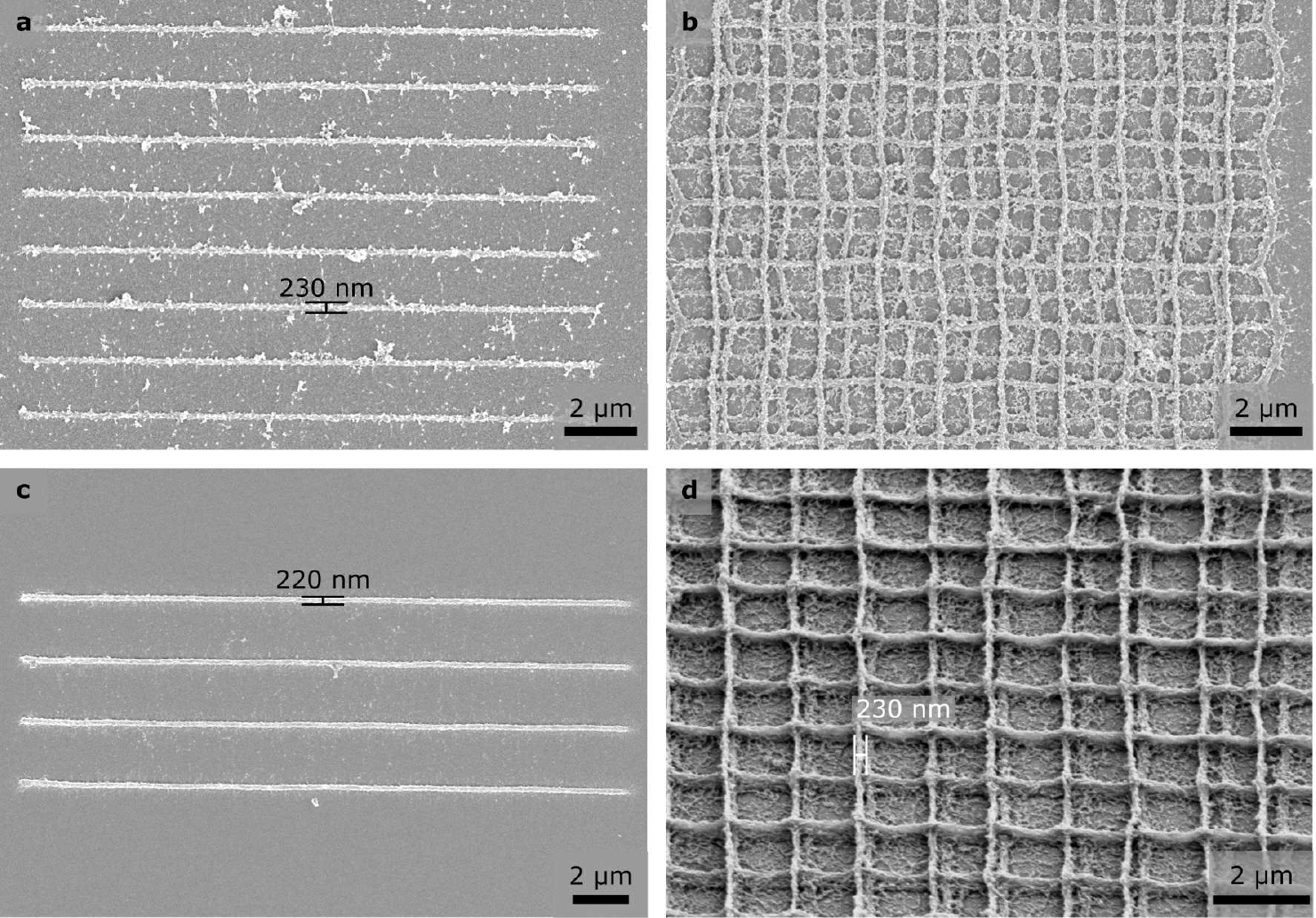}
    \caption{SEM micrographs of (a) direct laser written 2D lines and (b) four layer woodpile structure using a CDA-SH 4-pentenoate based resist. A writing speed of 100\,µm/s and 100\% laser power is used in both cases. (c) Printed 2D lines and (d) four layer grid structure applying a CDA-SH methacrylate based resist. A writing speed of 20\,µm/s is used in both cases as well as a laser power of 80\% (c) and 40\%  (d), respectively. (d) displays a side view at 45°.}
    \label{fig:DLW_structures}
\end{figure}

Figure \ref{fig:DLW_structures}a,b shows direct laser written structures in a resist possessing a concentration of 140\,g CDA-SH 4-pentenoate per liter DMF. Indeed, this reveals that introduced thiol functionalities allow for fabricating cellulose-based structures via a two-photon polymerization process without any additional photoinitiator. While two-dimensional structures (see Fig. \ref{fig:DLW_structures}a) with feature size down to 230\,nm can be readily fabricated, printing of three-dimensional architectures remains a challenge. Fig. \ref{fig:DLW_structures}b depicts a printed four layer woodpile structure revealing the general shape of a woodpile. However, the seemingly random formation of bulges at the written structures, which can be also seen in Fig. \ref{fig:DLW_structures}a, hinders an improved quality of the structure.

In contrast, the formation of bulges is almost absent in resists based on CDA-SH methacrylate as shown in Fig. \ref{fig:DLW_structures}c,d. Compared to Fig. \ref{fig:DLW_structures}a the two-dimensional lines in Fig. \ref{fig:DLW_structures}c are well defined and significantly sharper while a similar feature size of around 220\,nm is achieved. Thus, the CDA-SH methacrylate derivative clearly outperforms the CDA-SH 4-pentenoate in terms of two-dimensional structure fabrication. However, printing of complex three-dimensional structures is also challenging using CDA-SH methacrylate since the resist tends to form small, unintended filaments between features in close proximity. Therefore, complex structures tend to become overgrown with a polymer film inhibiting a good structure quality for complex architectures. Nevertheless, it is still possible to fabricate less complex structures such as the printed four layer grid structure displayed in Fig. \ref{fig:DLW_structures}d. Here, it can be discerned that some lines already bridging the underlying lines indicating the general potential for three-dimensional fabrication.

As in the case of two-dimensional structures also for simple three-dimensional structures line widths of around 230\,nm are obtained. This remarkably small feature size is only about the half of the feature size, which was state-of-the-art for cellulose-based resists up to now \cite{Rothammer2018}. Thus, fabrication of features on the order of half the wavelength of visible light become available with the here presented resists. 

The ability of printing such small features is highly needed for biomimicries of photonic structures found in nature, e.g. structural colors. Here, tailored disorder plays in many cases a crucial role to achieve the corresponding effect. For example the brilliant whiteness of ultrathin beetle scales \cite{Vukusic2007} stems from efficient light scattering within a complex intra-scale network featuring struts with a varying diameters in the range between 100\,nm and 600\,nm \cite{Wilts2018}. Indeed, it was demonstrated that this thickness variation on a very small length scale is responsible for the extraordinary whiteness \cite{Meiers2018}. However, in case of brilliant white structures the effect can be scaled to allow for micro-printing of architectures with comparable whiteness but on the cost of an increased overall structure thickness \cite{Pompe2022}. The situation is different if colored structures are considered such as found in several blue butterflies \cite{Kinoshita2005}. Here, the size of the features as well as a precise control of the disorder matters to achieve the coloration \cite{Watanabe2004}. Therefore, resists especially based on biomaterials, which enable feature sizes below the wavelength of visible light are still highly required to create more realistic copies of natural photonic structures possessing tailored disorder. Achieving such small feature sizes, the here presented derivatives, especially CDA-SH methacrylate, have high potential to serve as such resists although some further improvements has to be made to obtain more complex three-dimensional structures.

\section{Conclusion}
The class of bio-based photoresists applicable for crosslinking via DLW was expanded by bifunctional cellulose diacetates, which exhibit thiol moieties as well as olefinic or methacrylic side groups. Furthermore, these new class of biopolymers can be structured by two-photon absorption without the addition of a photoinitiator. Beyond that, these cellulose derivatives can be photo-crosslinked also via one-photon absorption employing a UV lamp without the need to apply an initiator. In case of writing structures via DLW remarkably small feature sizes well below the wavelength of visible light are achieved for two- and simple three-dimensional structures. While the fabrication of more complex three-dimensional architectures requires some further improvement, the here presented synthesis of bifunctional cellulose derivatives is an important step towards bio-based photoresists, which enable a close biomimicry of ordered and disordered architectures found in nature.

\medskip
\textbf{Funding} \par
The financial support from German Science Foundation (DFG) for funding this work within the priority program “Tailored Disorder – A science- and engineering-based approach to materials design for advanced photonic applications” (SPP-1839) is gratefully acknowledged. 

\medskip
\textbf{Acknowledgments} \par
Maximilian Rothammer and Dominic T. Meiers have equally contributed to this work. 

\medskip
\textbf{Disclosures} \par
The authors declare no conflicts of interest.

\medskip
\textbf{Supplemental document} \par
See supplemental document for supporting content.

\section{TOC}

\begin{figure}[th]
    \centering\includegraphics[width=15cm]{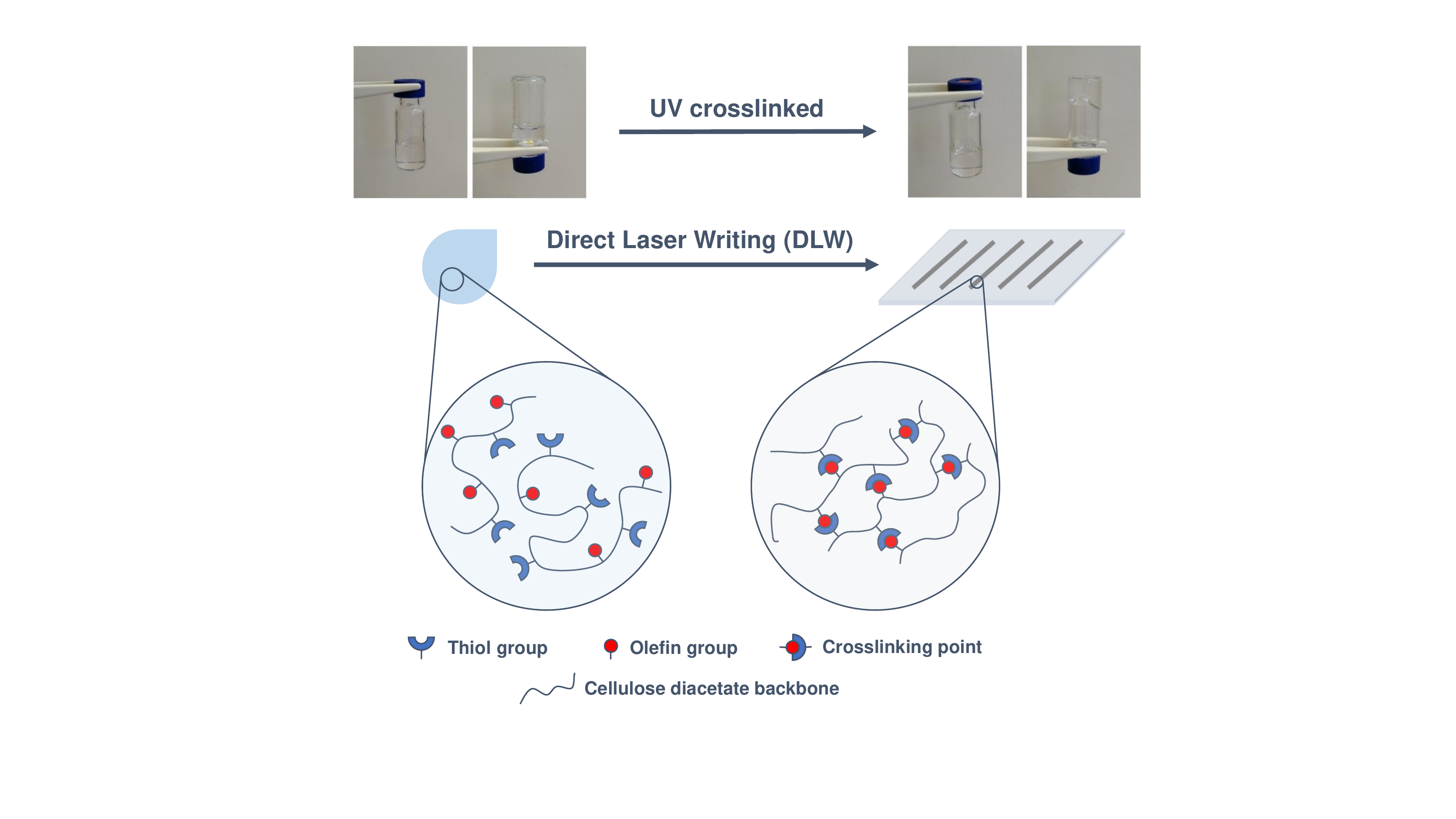}
    \caption{Table of contents}
    \label{fig:TOC}
\end{figure}

Novel bifunctional cellulose diacetate derivatives were synthesized to obtain a bio-based photoresist. This cellulose-based resist can be crosslinked by two-photon absorption via direct laser writing as well as by one-photon absorption via UV irradiation ($\lambda$ = 254\,nm and 365\,nm). Thereby, these photo-reactive biopolymers can be crosslinked without a photoinitiator. 

%%%%%%%%%%%%%%%%%%%%%%% References %%%%%%%%%%%%%%%%%%%%%%%%%

\newpage



\section*{Supplementary material}

# Initiator-free photo-crosslinkable cellulose-based resists for fabricating submicron patterns via direct laser writing: supplemental document

## 1. GENERAL INFORMATION

Unless otherwise stated, all chemicals were purchased from Sigma-Aldrich, VWR Chemicals, Alfa Aesar or Roth and used without further purification. The reductive disulfide cleavage was carried out under argon (99.996 vol.%) from Westfalen as inert gas.

## 2. SYNTHESIS PROCEDURES

**Synthesis Disulfide Compound**

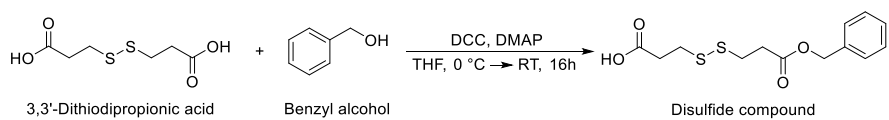

The synthesis of the disulfide compound was adapted from a literature-known procedure [1]. 3,3′-Dithiodipropionic acid (20.0 g, 95.1 mmol) was dissolved in 400 mL THF in a flask equipped with a magnetic stirrer. After benzyl alcohol (8.60 g, 79.2 mmol) and DMAP (1.00 g, 8.19 mmol) were added, the mixture was cooled to 0 °C. DCC (16.4 g, 79.2 mmol) was dissolved in 100 mL THF and added slowly via a dropping funnel. The reaction was allowed to proceed for 16 hours. After filtration of the crude reaction mixture, 100 mL of distilled water were added and the mixture was allowed to stir for 1 hour. The solvent was evaporated and 300 mL of DCM were added to the mixture. The organic phase was washed five times with 50 mL aqueous $KHSO_4$ (10 wt%) and dried over $MgSO_4$ and the solvent was evaporated. The product was recrystallized from DCM:Heptane 1:5 and obtained as a white solid. (55%, 13.18 g, 43.9 mmol)

$^1$H-NMR (400 MHz, $CDCl_3$): $\delta$ (ppm) = 7.36 (s, 5H), 5.15 (s, 2H), 2.98 – 2.87 (m, 4H), 2.82 – 2.74 (m, 4H).

$^{13}$C-NMR (101 MHz, $CDCl_3$): $\delta$ (ppm) = 177.25, 171.69, 135.77, 128.74, 128.46, 66.82, 34.27, 33.95, 33.20, 32.87.

**Disulfide Protection of CDA**

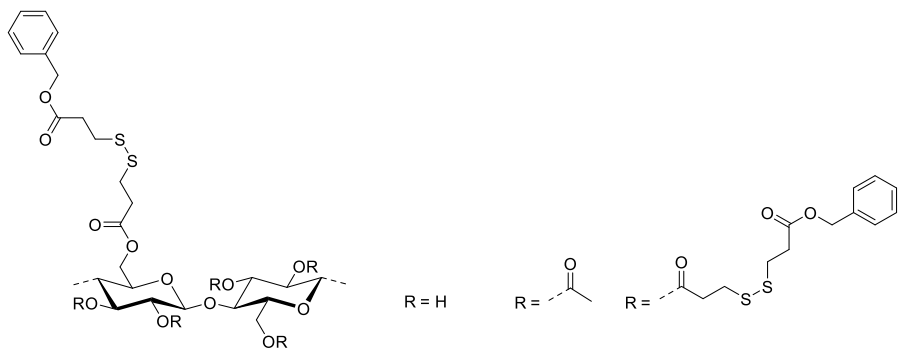

CDA (1.00 g, 3.76 mmol referred to AGU) was dissolved in 40 mL DCM in a flask equipped with a magnetic stirrer. DCC (0.87 g, 4.17 mmol), DMAP (0.02 g, 0.16 mmol) and the disulfide compound (1.21 g mmol, 4.17 mmol) were added to the reaction mixture. The reaction was performed under permanent stirring for 72 h. The modified CDA was precipitated in an excess of ethanol and was washed five times with ethanol. The product was dried under vacuum and was obtained as white solid (1.11 g, 91%).

$^{13}$C-NMR (101 MHz, CDCl$_3$): δ (ppm) = 171.4, 170.1, 169.3, 168.9, 136.5, 128.5 – 128.1, 100.4 – 62.4, 66.0, 33.8 – 32.9, 20.0

**Functionalization of dp-CDA with methacrylic anhydride or 4-pentenoic acid**

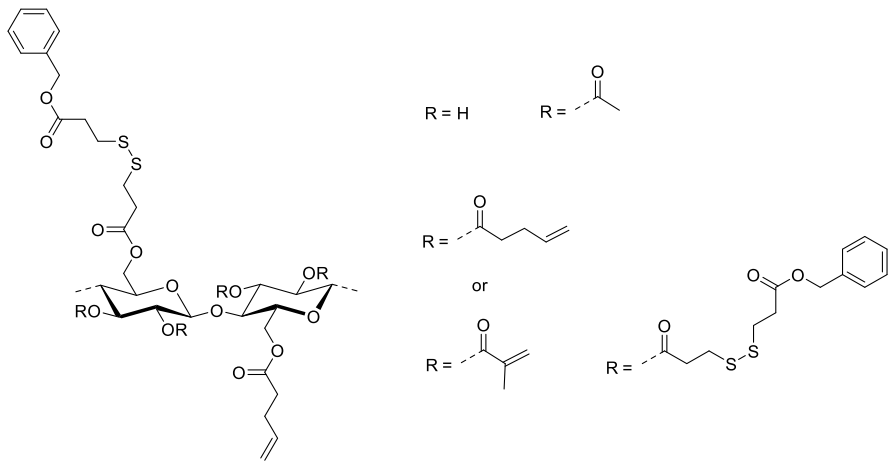

Dp-CDA (0.50 g) was dissolved in 20 mL DCM in a flask equipped with a magnetic stirrer. DCC (0.85 g, 4.12 mmol), DMAP (0.01 g, 0.08 mmol) and the olefin functional compound (5.70 mmol for methacrylic anhydride, 12.5 mmol for 4-pentenoic acid) were added to the reaction mixture. The reaction was performed under permanent stirring for 72 h. The modified CDA was precipitated in an excess of ethanol and was washed five times with ethanol. The product was dried under vacuum and was obtained as white solid.

dp-CDA methacrylate:
Yield: 92%
$^{13}$C-NMR (101 MHz, CDCl$_3$): δ (ppm) = 172.1, 171.0, 170.2, 169.8, 167.3, 137.5, 137.2, 126.9, 129.4 – 129.2, 101.3 – 63.2, 66.9, 34.0, 30.9, 18.5

dp-CDA 4-pentenoate
Yield: 91%
$^{13}$C-NMR (101 MHz, CDCl$_3$): δ (ppm) = 172.1, 171.9, 171.0, 170.2, 169.8, 137.9, 137.5, 129.1, 116.3, 101.3- 63.2, 66.4, 33.9 – 33.6, 32.7, 28.8, 20.6

**Reductive disulfide cleavage and Thiol-activation**

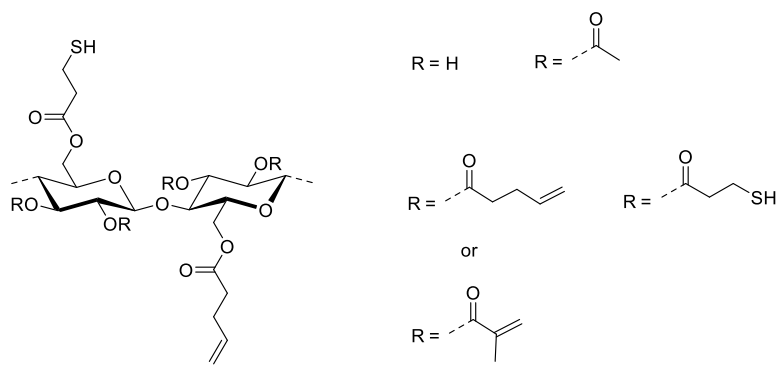

Functionalized dp-CDA (0.38 g for dp-CDA methacrylate, 0.35 g for dp-CDA 4-pentenoate) was dissolved in 20 mL CHCl$_3$ in a brown round bottom flask equipped with a magnetic stirrer. The flask was flushed with argon and DTT (0.21 g, 1.36 mmol) and TEA (0.40 mL, 2.89 mmol) were added to the reaction mixture. The flask was filled with argon and sealed with a septum. The reaction was allowed to proceed for 16 hours. After the reaction, the mixture was poured in an excess of ethanol and washed three times with ethanol. The product was dried under vacuum and was obtained as white solid. The product was stored in the dark at -20 °C.

CDA-SH methacrylate
Yield: 78%
$^{13}$C-NMR (101 MHz, CDCl$_3$): δ (ppm) = 171.9, 170.9, 170.1, 169.7, 167.1, 137.1, 126.9, 101.1 – 63.1, 38.9, 20.7, 20.1, 18.4

CDA-SH 4-pentenoate
Yield: 83%
$^{13}$C-NMR (101 MHz, CDCl$_3$): δ (ppm) = 171.9, 170.9, 170.1, 169.7, 137.8, 115.9, 101.2 – 63.1, 39.0, 33.9, 28.4, 20.6, 20.1

**REFERENCES**

1. O. Andrén and M. Malkoch, "Facile thiolation of hydroxyl functional polymers," Polym. Chem. **8**, 4996–5001 (2017).



\end{document}